\documentclass[aps,preprint,nofootinbib,preprintnumbers,eqsecnum]{revtex4-1}

\usepackage{color}

\newcommand{\nb}[1]{\color{blue}}

\newcommand{\hl}[1]{\color{magenta}}

\newcommand{\tr}{\mathrm{Tr}\,}

\newcommand{\arctanh}{\mathrm{arctanh}}
\newcommand{\eqn}[1]{\begin{equation}{\label{#1}}}
\newcommand{\eqq}{\end{equation}}
\newcommand{\beq}{\begin{equation}}

\usepackage[
	  pagebackref=false,
	  colorlinks=true,
      linkcolor=blue,
      urlcolor=blue,
      filecolor=black,
      citecolor=red,
      pdfstartview=FitV,
      pdftitle={},
        pdfauthor={Paolo Glorioso},
        pdfsubject={},
        pdfkeywords={},
        pdfpagemode=None,
        bookmarksopen=true
      ]{hyperref}

\usepackage[normalem]{ulem}
\usepackage{amsmath}
\usepackage{enumerate}
\usepackage{amsfonts}
\usepackage{epsfig}
\usepackage{array}

\def\tr{\mathop{\rm tr}}

\newcommand\p{\ensuremath{\partial}}

\newcommand{\be}{\begin{equation}}
\newcommand{\ee}{\end{equation}}
\newcommand{\bea}{\begin{eqnarray}}
\newcommand{\eea}{\end{eqnarray}}

\newcommand{\bi}{\begin{itemize}}
\newcommand{\ei}{\end{itemize}}
\newcommand{\ben}{\begin{enumerate}}
\newcommand{\een}{\end{enumerate}}
\newcommand{\bca}{\begin{cases}}
\newcommand{\eca}{\end{cases}}
\newcommand{\bln}{\begin{align}}
\newcommand{\eln}{\end{align}}
\newcommand{\bst}{\begin{split}}
\newcommand{\est}{\end{split}}

\begin{document}
\rightline{MIT-CTP/4736}
\title{Classification of certain asymptotically AdS space-times with Ricci-flat boundary}


\author{Paolo Glorioso}
\affiliation{Center for Theoretical Physics, \\
Massachusetts
Institute of Technology,
Cambridge, MA 02139 }

\begin{abstract}

\noindent
We classify solutions to Einstein's equations in AdS with Ricci-flat boundary metric and with covariantly constant boundary stress tensor, which in general is not diagonalizable, i.e. it does not admit a reference frame. New solutions are found, and in the context of the AdS/CFT duality they should describe a boundary QFT in certain non-equilibrium steady states. Further imposing the absence of scalar curvature singularities leads to a subset of metrics that can be seen as null deformations of AdS or of the AdS soliton. 
We also outline the procedure of solving the equations when a scalar is coupled to the metric, which holographically leads to non-Lorentz-invariant RG flows.

\end{abstract}

\today

\maketitle

\tableofcontents

\section{Introduction}
The gauge/gravity duality enables us to explore a wide range of aspects of quantum field theories in the strongly coupled regime \cite 6. The universal sector of the dynamics on the string theory side in the semiclassical limit is given by Einstein's equation with negative cosmological constant. It is therefore of great interest to classify the solutions to this equation, and explore their structure.

In this paper we classify the most general solution to Einstein's equation with negative cosmological constant subject to the condition that the metric of the dual QFT is Ricci-flat, and the stress tensor is covariantly constant. Our main focus is the case of flat boundary space-time, in which case the classification is solely determined by the form of the stress tensor, for which in general there is no rest frame, i.e. it cannot be put in diagonal form. Such solutions should correspond to certain non equilibrium steady states, and should be useful for uncovering new non-equilibrium phenomena.

In the next Section we present the most general solution to Einstein's equation in AdS gravity with Ricci-flat boundary metric and covariantly constant boundary stress tensor. In Section 3 we focus on solutions with flat boundary space-time. In this case, the boundary stress tensor admits four algebraically distinct forms, in the sense that they cannot be related to each other by Lorentz transformations. This classification is naturally extended to the corresponding gravity solutions. Next, it is shown that requiring the absence of scalar curvature singularities leads to metrics that can be interpreted as ``null deformations'' of AdS or of the AdS soliton \cite {7,18a}.

In Section 4 we extend our classification to solutions with Ricci-flat boundary metric. Here the classification is given in terms of the holonomy group of the boundary metric and of the form of the boundary stress tensor. These metrics can be thought of as dual to out of equilibrium steady states on a Ricci-flat space-time.

In Section 5 we extend the above setup to gravity coupled to a scalar. We show that the system is equivalent to a domain wall coupled to two scalars, and implement on it a well-known method \cite 3, which simplifies further the structure of the equations of motion by introducing a fictitious superpotential. On the QFT side, the solutions to these equations can be interpreted as RG flows that do not preserve Lorentz symmetry. In Section 6 we conclude with a discussion of our results and the outlook.

\section{General solution}\label{gensol}
The Einstein-Hilbert action with cosmological constant is
\be\label{actme} S=\int_M d^{d+1} x\sqrt{-G}\left(R-2\Lambda\right)-2\int_{\p M}d^d x\sqrt{-\gamma}K+S_{\text{ct}},\ee
where $M$ is a $(d+1)$-dimensional space-time with boundary $\p M$, and
$\Lambda=-d(d-1)/2<0$ is the cosmological constant.\footnote{We use the convention
$R^M_{NAB}=\partial_A\Gamma^M_{NB}-\partial_B\Gamma^M_{NA}
+\Gamma^M_{CA}\Gamma^C_{NB}-\Gamma^M_{CB}\Gamma^C_{NA}$.} The second integral is
the Gibbons-Hawking term, which is included to guarantee a well-defined variational principle, where $K=K^M_{\ M}$ is the trace of the extrinsic curvature
$K_{MN}=\frac12 \mathcal L_n G_{MN}$, where $n$ is the unit normal of $\p M$.
$S_{\text{ct}}$ depends only on intrinsic quantities defined on $\p M$,
it is included to guarantee that the boundary stress tensor defined below is finite, and it has the form
\be\label{actgh} S_{\text{ct}}=-2(d-1)\int_{\p M}d^d x\sqrt{-\gamma}+\cdots,\ee
where the dots stand for higher derivative terms. From the action (\ref{actme}) one obtains the Einstein equations
\eqn{einsteq} R_{MN}-\frac 12 G_{MN}R=-\Lambda G_{MN}.\eqq
We choose the coordinate system $x^M=(u,x^\mu)$ such that the metric takes the form
\eqn{gnc11} ds^2=G_{MN}dx^M dx^N=du^2+\gamma_{\mu\nu}dx^\mu dx^\nu,\eqq
and, for solutions to the above equation one has the behavior \cite{10}
\eqn{bdry1} \gamma_{\mu\nu}(u,x)\to e^{-2u}g_{\mu\nu}(x),\eqq
as $u\to -\infty$, which is where we shall place the boundary $\p M$. In (\ref{bdry1}), $g_{\mu\nu}$ is a $d$-dimensional metric which we take to have Lorentzian signature. The boundary stress tensor is defined as \cite{bala2}
\eqn{bst1} T^{\mu\nu}=\frac 2{\sqrt{-g}}\frac{\delta S}{\delta g_{\mu\nu}}=\lim_{u_c\to -\infty}2e^{-(d+2)u_c}\left(K^{\mu\nu}-K\gamma^{\mu \nu}-(d-1)\gamma^{\mu\nu}\right),\eqq
where in the last expression we neglected higher derivative terms coming from (\ref{actgh}), and where $K_{\mu\nu}$ is the extrinsic curvature of the hypersurface $u_c=const.$, which in the coordinate system (\ref{gnc11}) reads
\beq K^\mu_{\ \nu}=\frac 12\gamma^{\mu\alpha} \partial_u\gamma_{\alpha\nu}.\eqq
Suppose now that $g_{\mu\nu}$ is a Ricci-flat metric, i.e. its Ricci tensor vanishes, and that the boundary stress tensor (\ref{bst1}) is covariantly constant with respect to the connection of $g_{\mu\nu}$, i.e.
\eqn{covconst} \nabla_\mu T_{\nu\rho}=0,\eqq
where $\nabla_\mu$ is the covariant derivative compatible with $g_{\mu\nu}$. Then the solution to (\ref{einsteq}) can be written in closed form and it reads, in matrix notation
\eqn{sol0} \gamma=ge^{-2u}\left(1+ e^{d(u-u_0)}\right)^{\frac 2d (1+2^{-2}e^{du_0}T)}\left(1-e^{d(u-u_0)}\right)^{\frac 2d (1-2^{-2}e^{du_0}T)},\eqq
where $u_0$ is a constant scalar, $T$ is the matrix that represents the boundary stress tensor, and is subject to
\eqn{uvsym}(gT)^t=gT,\quad\text{i.e.}\quad T_{\mu\nu}=T_{\nu\mu},\eqq
\eqn{sol10}\tr T=0\eqq
and
\eqn{sol20}\tr T^2=16d(d-1)e^{-2du_0}.\eqq
The above result is derived in Appendix \ref{resalb}. Taking $u_0\to \infty$ we obtain an interesting limit case of the above solution, and it reads
\eqn{sol0a} \gamma=ge^{-2u}e^{\frac 1d e^{du}T},\eqq
with
\eqn{sol0a2} \tr T=0\,\quad \tr T^2=0.\eqq
The way (\ref{sol0}) and (\ref{sol0a}) are written is not particularly illuminating. In the following two Sections we shall give a classification of the various forms that these metrics can acquire. For each form we will be able to write the metric in a more transparent way and study various aspects of it.

\section{Homogeneous solutions}\label{classification}
In this Section we analyze the case in which the boundary metric $g_{\mu\nu}$ defined in (\ref{bdry1}) is flat. Up to a coordinate change, we can then take
\eqn{guv1} g_{\mu\nu}=\eta_{\mu\nu},\eqq
where $\eta_{\mu\nu}=\mathrm{diag}(-1,1,\dots,1)$. We will first show that there is a natural classification of the forms that the boundary stress tensor can acquire. Such classification will be inherited by the corresponding gravity solutions.

\subsection{Canonical forms of the boundary tensor}\label{canby}
In Appendix (\ref{appendixa}) we show that the bulk coordinate transformations that preserve the form (\ref{gnc11}), the boundary condition (\ref{guv1}), and that keep the metric independent of $x^\mu$ are given by the Lorentz transformations acting on the $x^\mu$ coordinates and by constant translations in the $u$-direction. Such transformations act as Lorentz transformations and rescalings of $T^\mu_{\ \nu}$. It is then natural to introduce a classification of ``canonical forms'' of $T^\mu_{\ \nu}$ with respect to such transformations, i.e. representative forms that cannot be related to each other.

With the choice (\ref{guv1}), eq. (\ref{uvsym}) becomes
\eqn{lorsym} (\eta T)^t=(\eta T),\eqq
which tells that $T^\mu_{\ \nu}$ is not a symmetric matrix, and it turns out to be in general non-diagonalizable . In this case there are four canonical forms that $T^\mu_{\ \nu}$ can acquire, which can be written as
\eqn{matrixt1} T=\begin{pmatrix}S&\\&D\end{pmatrix},\eqq
where $D=\text{diag}(p_i,\dots,p_{d-1})$ is a diagonal $(d-3)$-dimensional matrix, with $i=3,\dots,d-1$, $d$ is the number of space-time dimensions, and $S$ is a $3\times 3$ matrix that has one of the following forms:
\begin{eqnarray}
\label{first} &\begin{pmatrix}p_0&&\\&p_1&\\&&p_2\end{pmatrix}\quad&\rightarrow\quad T^{(I)}_{\mu\nu}=-p_0 \delta^0_\mu \delta^0_\nu+p_1 \delta^1_\mu \delta^1_\nu+p_2 \delta^2_\mu \delta^2_\mu+p_i \delta^i_{\mu} \delta^i_{\nu}\\
\label{second} &\begin{pmatrix}p_0-\gamma & \gamma&\\ -\gamma& p_0+\gamma &\\&&p_2\end{pmatrix}\quad&\rightarrow\quad T^{(II)}_{\mu\nu}= -p_0 \delta^0_\mu \delta^0_\nu+p_0 \delta^1_\mu \delta^1_\nu+p_2 \delta^2_\mu \delta^2_\nu+\gamma l_\mu l_\nu+p_i \delta^i_{\mu} \delta^i_{\nu}\\
\label{third} &\begin{pmatrix}p_0&0&1\\0& p_0&1\\-1&1&p_0\end{pmatrix}\quad&\rightarrow\quad T^{(III)}_{\mu\nu}= -p_0 \delta^0_\mu \delta^0_\nu+p_0 \delta^1_\mu \delta^1_\nu+p_0 \delta^2_\mu \delta^2_\mu+2 l_{(\mu} \delta^2_{\nu)}+p_i \delta^i_{\mu} \delta^i_{\nu}\\
\label{fourth} &\begin{pmatrix}\rho_0&\rho_1&\\-\rho_1&\rho_0&\\&&p_2\end{pmatrix}\quad&\rightarrow\quad T^{(IV)}_{\mu\nu}= -\rho_0 \delta^0_\mu \delta^0_\nu+\rho_0 \delta^1_\mu \delta^1_\nu+p_2 \delta^2_\mu \delta^2_\nu-2\rho_1\delta^0_{(\mu}\delta^1_{\nu)}+p_i \delta^i_{\mu} \delta^i_{\nu}\qquad
\end{eqnarray}
where on the right we put the expression for the $d$-dimensional $T_{\mu\nu}$ in components, with $l^\mu=\delta_0^\mu+\delta_1^\mu$, $l^2=0$.

The form (\ref{second}) of the matrix block $S$ has two eigenvectors, with eigenvalues $p_0$ and $p_2$, and $\gamma$ can be set to $\gamma=\pm 1$ performing a boost along $x^1$. When $S$ has the form (\ref{third}) there is only one eigenvector, with eigenvalue $p_0$, and when $S$ has the form (\ref{fourth}) it has three eigenvectors, with eigenvalues $\rho_0\pm i\rho_1$ and $p_2$. Note in particular that the forms (\ref{second}) and (\ref{third}) are non-diagonalizable, and can be written in the form
\eqn{t23} T^{(II),(III)}_{\mu\nu}=D_{\mu\nu}+H_{\mu\nu},\eqq
where $D^{\mu}_{\ \nu}$ is diagonalizable, and $H^\mu_{\ \nu}$ is nilpotent, i.e. $H^n=0$ for some integer $n$, and $[D,H]=0$. A derivation of the above classification can be found in \cite 5. This classification is also given in \cite{12}, and we shall adopt the same nomenclature here: the forms (\ref{first})-(\ref{fourth}) are called type I, II, III and IV, respectively.

Note that with Euclidean metric $g_{\mu\nu}=\mathrm{diag}(1,\dots,1)$ this classification would be trivial. Condition (\ref{uvsym}) would imply that $T^\mu_{\ \nu}$ is a symmetric matrix, which can then be always diagonalized, and (\ref{first}) would be the only canonical form.

\subsection{Gravity solutions of type I and type IV}
We now turn to explore the gravity solutions and study their relation to the classification we gave above. In the reminder of the paper, we shall call metric of type I, II, III or IV the solutions (\ref{sol0}) obtained from type I, II, III or IV boundary stress tensors, respectively. In Appendix (\ref{curvsing}) we show that the solution (\ref{sol0}) has a scalar curvature singularity unless all but one eigenvalues of the boundary stress tensor $T_{\mu\nu}$ are equal to $4e^{-du_0}$.\footnote{By eigenvalues of $T_{\mu\nu}$ we mean the eigenvalues of the matrix $T^\mu_{\ \nu}$.} The remaining eigenvalue is determined by (\ref{sol10}), so we have:\footnote{The fact that having multiple different eigenvalues of the boundary stress tensor leads to singular solutions was known already, see e.g. \cite{16}.}
\eqn{eigT} p_i=\begin{cases}-4(d-1)e^{-du_0} &\text{for exactly one given $i$}\\ 4e^{-du_0}&\text{otherwise}\end{cases}\eqq
where $i=0,\dots,d-1$. These eigenvalues are consistent with condition (\ref{sol20}). For type I metrics there are three inequivalent possible cases. Either $e^{-du_0}= 0$ or $e^{-du_0}\neq 0$, where the latter has two subcases: $p_0=-4(d-1)e^{-du_0}$ with $p_{i\neq 0}=4e^{-du_0}$, or $p_{1}=-4(d-1)e^{-du_0}$ with $p_{i\neq 1}=4e^{-du_0}$. In matrix notation,
\eqn{firstcl} T=0,\quad T=4e^{-du_0}\mathrm{diag}(1-d,1,\dots,1),\quad T=4e^{-du_0}\mathrm{diag}(1,1-d,1,\dots,1).\eqq
Plugging the first case in (\ref{sol0}) we find
\beq ds^2=du^2+e^{-2u}\left(-(dx^0)^2+(dx^i)^2\right),\eqq
which is AdS. Plugging the second case,
\beq ds^2=du^2+e^{-2u}\left(1+ e^{d(u-u_0)}\right)^{\frac 4d}\left(-\left(\frac{1-e^{d(u-u_0)}}{1+e^{d(u-u_0)}}\right)^2(dx^0)^2+(dx^i)^2\right),\eqq
which is the black-brane, and $u_0$ is the position of the horizon, and plugging in (\ref{sol0}) the third form in (\ref{firstcl}) gives
\beq ds^2=du^2+e^{-2u}\left(1+ e^{d(u-u_0)}\right)^{\frac 4d}\left(-(dx^0)^2+\left(\frac{1-e^{d(u-u_0)}}{1+e^{d(u-u_0)}}\right)^2(dx^{1})^2+(dx^i)^2\right),\eqq
which is the AdS soliton, where $u_0$ is the location of the termination of space-time. All the other nonsingular solutions in this class are obtained by acting with Lorentz transformations on the above solutions.

Comparing (\ref{eigT}) with the possible eigenvalues of (\ref{fourth}) one easily sees that all the solutions of type IV have scalar curvature singularities for $d>2$. For completeness, we quote the expression of these metrics:
\eqn{fourthmetric}\begin{split} ds^2=&du^2+e^{-2u}(1-e^{2d(u-u_0)})^{\frac2d}\\
&\times\left(e^{-\lambda_0}\left(-\cos \lambda_1 (dx^0)^2-2\sin\lambda_1dx^0 dx^1+\cos\lambda_1(dx^1)^2\right)+e^{-\lambda_i}(dx^i)^2\right),\end{split}\eqq
where $i=2,\dots,d$, and
\beq\lambda_0=\frac 1{2d} e^{du_0}\rho_0 \log\left(\frac{1-e^{d(u-u_0)}}{1+e^{d(u-u_0)}}\right),\quad \lambda_1=\frac 1{2d} e^{du_0}\rho_1 \log\left(\frac{1-e^{d(u-u_0)}}{1+e^{d(u-u_0)}}\right),\eqq
\beq\lambda_i=\frac 1{2d} e^{du_0}p_i \log\left(\frac{1-e^{d(u-u_0)}}{1+e^{d(u-u_0)}}\right),\eqq
where $\rho_0,\rho_1,p_i$ are defined in (\ref{fourth}), with $i=2,\dots,d-1$ and are subject to (\ref{sol10}) and (\ref{sol20}).

\subsection{Gravity solutions of type II and type III: null deformations}\label{ndsec}
As we argued in the previous part, in order to avoid scalar curvature singularities the eigenvalues of the stress tensor should be of the form given in (\ref{eigT}). Comparing this with (\ref{second}) and (\ref{third}) we find that type II and type III stress tensors of nonsingular solutions, up to Lorentz transformations, should have the form
\begin{eqnarray}
\label{stii} T^{(II)}_{\mu\nu}&=& 4e^{-d u_0}\left(- \delta^0_\mu \delta^0_\nu+ \delta^1_\mu \delta^1_\nu+\delta^2_\mu \delta^2_\nu-(d-1) \delta^3_\mu \delta^3_\nu+ \delta^i_{\mu} \delta^i_{\nu}\right)+\gamma l_\mu l_\nu\\
\label{stiii} T^{(III)}_{\mu\nu}&=& 4e^{-d u_0}\left( -\delta^0_\mu \delta^0_\nu+ \delta^1_\mu \delta^1_\nu+\delta^2_\mu \delta^2_\nu-(d-1) \delta^3_\mu \delta^3_\nu+ \delta^i_{\mu} \delta^i_{\nu}\right)+2 l_{(\mu} \delta^2_{\nu)},\end{eqnarray}
where $i=4,\dots,d-1$. In both (\ref{stii}) and (\ref{stiii}), the expression proportional to $e^{-du_0}$ is the stress tensor corresponding to the AdS soliton when $u_0$ is finite, and to AdS when $u_0\to\infty$. The bulk metrics corresponding to (\ref{stii}) and (\ref{stiii}) can be seen as ``null deformations'' of AdS or of the AdS soliton. Indeed, plugging (\ref{stii}) or (\ref{stiii}) in (\ref{sol0}) we obtain a metric of the form
\eqn{nuldef0} \gamma_{\mu\nu}=\bar \gamma_{\mu\nu}+f_1(u)l_{(\mu} \delta^2_{\nu)}+f_2(u) l_\mu l_\nu,\eqq
where $\bar \gamma_{\mu\nu}$ is either AdS or the AdS soliton, and $f_1,f_2$ are some functions of the radial coordinate $u$. These metrics differ from $\bar \gamma_{\mu\nu}$ by specific combinations involving the vector $l_\mu$, which is null with respect to the boundary metric $\eta_{\mu\nu}$. In this sense, the expression (\ref{nuldef0}) can be seen as a generalization of the null deformations of AdS studied e.g. in \cite{7,18a}. They are not a continuous deformations of $\bar \gamma_{\mu\nu}$, and they should not be seen as perturbations of the latter. Note also that it is not algebraically possible to have null deformations of the black brane, as the stress tensor corresponding to the latter is not compatible with the forms (\ref{second}) and (\ref{third}), having different eigenvalues in time and space directions. The null deformations arising from the general solution (\ref{sol0}) are schematically summarized in Table \ref{table1}.
\begin{table}[h]\begin{center}
\begin{tabular}{p{3cm}|p{3cm}|p{3cm} l}
\centering{Type I}&\centering{Type II}&\centering{Type III}&\\
\hline
\centering{AdS}&\centering{AdS II$^+$, AdS II$^-$}&\centering{AdS III}&\\
\hline
\centering{Sol, BB}&\centering{Sol II$^+$, Sol II$^-$}&\centering{Sol III}&\\
\end{tabular}
\caption[Table 1]{Gravity solutions with no scalar curvature singularities. Type II and III can be obtained by null deformations of type I metrics.} \label{table1} \end{center}\end{table}\vspace{-0.5cm}
\\The functions $f_1,f_2$ in (\ref{nuldef0}) depend on the dynamics of the bulk gravity theory. In the reminder of this Section we shall explore the explicit expression of (\ref{nuldef0}) in Einstein gravity.

As shown in Table \ref{table1}, in type II solutions we have two null deformations of AdS, as well as two null deformations of the AdS soliton. Plugging (\ref{stii}) with $u_0=\infty$ in (\ref{sol0a}) we find
\eqn{nametric} ds^2=du^2+e^{-2u}\left(-2dx^+dx^-+\frac 2d\gamma e^{du} (dx^-)^2+(dx_\perp)^2\right),\eqq
where $dx^{\pm}=\frac1{\sqrt 2}(dx^0\pm dx^1)$, $dx_\perp^2$ refers to the other $d-2$ coordinates, and $\gamma=\pm 1$. This metric is the prototype of AdS plane wave, which has been studied in particular when $\gamma>1$ \cite{7,18a}. We shall denote the above metric by AdS II$^+$ when $\gamma=1$, and by AdS II$^-$ when $\gamma=-1$. One can show that AdS II$^+$ can be obtained from the black brane in the infinite boost limit $dx^-\to \lambda dx^-$, $dx^+\to \lambda^{-1}dx^+$, with $\lambda\to\infty$, where horizon is pushed to the IR infinity so that $e^{-du_0}\lambda^2=\frac2d\gamma $ \cite{15}. Analogously, one can obtain  AdS II$^-$ from the soliton by taking the same limit (this time with $e^{-du_0}\lambda^2=-\frac2d\gamma $), although the geometric meaning of this procedure is not clear to us. In Appendix \ref{apptidal} we find that AdS II$^+$ exhibits tidal force singularities, which are detected by geodesics that reach the IR infinity in a finite amount of proper time. These singularities are of the same type as those found in Lifshitz space-times \cite{8,9,bala,Hart}. In AdS II$^-$ there are no geodesics that can reach the IR infinity in a finite proper time so that, in contrast to AdS II$^-$, no tidal force singularities are detected.

The other solution of type II that has no scalar curvature singularity can be obtained from the stress tensor $T^{(II)}_{\mu\nu}$ in (\ref{stii}) with $u_0$ finite. Plugging it in (\ref{sol0}) yields
\eqn{solit2} \begin{split}ds^2=&du^2+e^{-2u}\left(1+ e^{d(u-u_0)}\right)^{\frac 4d}\\
&\times\bigg(-2dx^+dx^-+\frac 2d\gamma \,\arctanh\left(e^{d(u-u_0)}\right)(dx^-)^2+\tanh^2\left(\frac d2(u-u_0)\right)(dx^2)^2+(dx_\perp)^2\bigg),\end{split}\eqq
which can be seen as a null deformation of the AdS soliton. We shall denote the above metric by Sol II$^+$ when $\gamma=1$, and Sol II$^-$ when $\gamma=-1$. As shown in Appendix \ref{apptidal}, there is a tidal force singularity at $u_0$, which for Sol II$^+$ can be reached by time-like geodesics, and for Sol II$^-$ can be reached only by space-like geodesics.

For type III solutions we have two null deformations, one of AdS and one of the soliton. Inserting (\ref{stiii}) with $u_0=\infty$ in (\ref{sol0a}) gives
\eqn{ads3} ds^2=du^2+e^{-2u}\left(-2dx^+dx^-+(dx^2)^2-\frac {2\sqrt 2}d e^{du} dx^-dx^2+\frac{1}{d^2} e^{2du}(dx^-)^2+(dx_\perp)^2\right),\eqq
which we shall denote by AdS III. The singularity structure is similar to that of AdS II$^-$: there are no geodesics that can reach the IR infinity in a proper time, and thus there is no detection of tidal force singularities.

Plugging (\ref{stiii}) with $u_0$ finite in (\ref{sol0a}) gives the type III null deformation of the soliton,
\eqn{solit3}\begin{split} ds^2=&du^2+e^{-2u}\left(1+ e^{d(u-u_0)}\right)^{\frac 4d}\bigg(-2dx^+dx^-+(dx^2)^2+\tanh^2\left(\frac d2(u-u_0)\right)(dx^3)^2\\
&-\sqrt 2\frac 2d \arctanh\left(e^{d(u-u_0)}\right)dx^-dx^2+\frac{1}{d^2}\arctanh^2\left(e^{d(u-u_0)}\right)(dx^-)^2+(dx_\perp)^2\bigg),
\end{split}\eqq
which we denote by Sol III. As shown in Appendix \ref{apptidal}, this solution has a tidal force singularity at $u_0$ which, like Sol II$^-$, can be reached only by space-like geodesics. We note that we were not able to find the above metric in the literature, as well as (\ref{solit2}) and (\ref{ads3}).\footnote{For type I and II solutions our results agree with \cite{Bhaseen:2013ypa}, where boosted black branes were shown to be the only regular solutions with a homogeneous stress tensor with non-negative energy density. However, also some type III solutions here appear to satisfy such properties. More specifically, the AdS III metric (\ref{ads3}) is regular and has a stress tensor with non-negative energy density, given by (\ref{stiii}) with $u_0=\infty$. We thank Julian Sonner and the authors of \cite{Bhaseen:2013ypa} for the helpful discussions regarding this.}

\section{Non-homogeneous solutions}\label{nonhomsec}
In this Section we explore solutions for which the metric $g_{\mu\nu}$ defined in (\ref{bdry1}) is Ricci-flat but not flat. Given $g_{\mu\nu}$, a specific solution is determined by plugging in (\ref{sol0}) a boundary stress tensor that satisfies eq. (\ref{covconst}), i.e. it is covariantly constant.  Such tensors are subject to an algebraic classification that depends on the holonomy group of the space-time of $g_{\mu\nu}$. The holonomy groups of simply connected space-times are Lie subgroups of the Lorentz group, and therefore they can be described by the Lie subalgebras of the Lorentz algebra. We shall restrict to boundary dimension $d=4$. The holonomy groups of simply connected Ricci-flat space-times are generated by the Lorentz subalgebras $R_1$, $R_8$, $R_{14}$ and $R_{15}$ \cite 5, which are introduced in Appendix \ref{holap}.\\
The holonomy type $R_1$ corresponds to flat space-time, which we analyzed in the previous Section. The holonomy types $R_{14}$ and $R_{15}$ admit only covariantly constant symmetric tensors proportional to the metric, $T_{\mu\nu}=\lambda g_{\mu\nu}$, where  $\lambda$ is a constant. Eq. (\ref{sol10}) implies then $T=0$, and therefore the only possible solution is
\begin{align} ds^2=du^2+e^{-2u}g_{\mu\nu}dx^\mu dx^\nu,\end{align}
which has the same bulk profile of AdS but with a non-flat $g_{\mu\nu}$. The holonomy type $R_8$ is more interesting. In this case we can have
\eqn{r8} T_{\mu\nu}=\lambda g_{\mu\nu}+\gamma \ell_\mu \ell_\nu,\eqq
where $\lambda$ and $\gamma$ are constants, and $l_\mu$ is a covariantly constant null vector, i.e.
\beq \nabla_\mu \ell_\nu=0,\eqq
with $l^2=0$. Compatibility with (\ref{sol10}) requires that $\lambda=0$ in (\ref{r8}), so that
\eqn{r8a} T_{\mu\nu}=\gamma \ell_\mu \ell_\nu,\eqq
which satisfies the second eq. in (\ref{sol0a2}). Plugging in (\ref{sol0a}), we find the bulk metric
\eqn{r8bulk}ds^2=du^2+e^{-2u}(g_{\mu\nu}+ \gamma e^{4u} \ell_\mu \ell_\nu)dx^\mu dx^\nu,\eqq
which can be seen as a generalization of (\ref{nametric}). One well-known example of type $R_8$ metric is the pp-wave space-time
\eqn{ppw} g_{\mu\nu}dx^\mu dx^\nu=H(x^-,x^2,x^3)d(x^-)^2+2dx^+dx^-+(dx^2)^2+(dx^3)^2,\eqq
with $\partial_2^2H+\partial_3^2 H=0$, and $\ell_\mu=\partial_\mu x^-$. The total bulk metric then reads
\beq ds^2=du^2+e^{-2u}\left((H(x^-,x^2,x^3)+ \gamma e^{4u})d(x^-)^2+2dx^+dx^-+(dx^2)^2+(dx^3)^2\right).\eqq
We end this Section by mentioning that any Ricci-flat space-time with nowhere vanishing Riemann tensor and admitting a nowhere zero covariantly constant vector $\ell_\mu$ is locally isometric to (\ref{ppw}) \cite{5}.

\section{Coupling to scalars}\label{coupsc}
The resolution technique that we applied to find (\ref{sol0}) can be extended to the case in which scalars are coupled to gravity. For simplicity we shall consider coupling to one scalar only, in which case the action is given by
\eqn{initialaction} S=\int_M d^{d+1}x \sqrt{-G}\left( R-\frac 12 (\partial_M\phi)^2-V(\phi)\right)-2\int_{\partial M}\sqrt{-\gamma} K+S_{\text{ct}},\eqq
where $S_{\text{ct}}$ is the counterterm action, and $V(\phi)$ is the scalar potential. In order to have the correct asymptotics, we assume that $\phi$ approaches a critical point $\bar\phi$ of $V$ as $u\to-\infty$. More explicitly, taking
\be V(\phi)=V(\bar\phi)+\frac 12 m^2 (\phi-\bar \phi)^2+\cdots,\ee
the behavior of $\phi$ near the boundary is given by
\be\label{phia} \phi(u,x^\mu)\to \bar\phi+\phi_{(0)}(x^\mu) e^{(d-\Delta)u},\ee
where $\Delta=\frac d2+\frac 12\sqrt{d^2+4m^2}$, and where we take $m^2<0$ so that that the second term in (\ref{phia}) vanishes as $u\to -\infty$. Note that we also require $m^2$ to satisfy the BF bound, $m^2>-\frac{d^2}4$.\footnote{We refer the reader to \cite{DHoker:2002nbb} for more details on this.} In order to cancel the divergent terms in (\ref{initialaction}), we then take
\be\label{sctsc} S_{\text{ct}}=-\int_{\p M}d^d x\sqrt{-\gamma}\left(2(d-1)+\frac {d-\Delta}2(\phi-\bar\phi)^2+\cdots\right),\ee
where the first term is the gravity contribution (\ref{actgh}), and the dots stand for higher derivative terms. Finally, we introduce the VEV of the boundary operator $\mathcal O$ conjugate to $\phi_{(0)}$
\be \mathcal O=\frac{1}{\sqrt{-g}}\frac{\delta S}{\delta\phi_{(0)}}=(\phi'-(d-\Delta)\phi+\cdots)e^{-\Delta u},\ee
where in the last expression we neglected higher derivative terms coming from (\ref{sctsc}).

In the reminder of this Section we outline an approach to find solutions to the bulk equations associated to (\ref{initialaction}) such that, as before, the boundary metric is Ricci flat and the stress tensor is covariantly constant. Additionally, we require that $\phi_{(0)}$ and $\mathcal O$ satisfy\footnote{As a consequence of diffeomorphism invariance in the bulk, $T^{\mu\nu}$ and $\mathcal O$ satisfy the Ward identity \cite{Bianchi:2001kw}
\be \nabla_\mu T^{\mu\nu}=\mathcal O\p^\nu\phi_{(0)}.
\ee
In particular, requiring $T^{\mu\nu}$ to be covariantly constant implies already $\p_\mu \phi_{(0)}=0$.
}
\be\label{const1} \p_\mu\phi_{(0)}=\p_\mu\mathcal O=0.\ee
Following the lines of Appendix \ref{resalb}, we assume (\ref{ricciflatu}), together with $\p_\mu\phi=0$ across the whole bulk. We also assume that the extrinsic curvature defined in (A4) is everywhere covariantly constant.\footnote{In Appendix \ref{resalb} we did not make this assumption as we could imply it from the explicit form of the solution (\ref{solk1}),(\ref{constraints1}), together with condition (\ref{covconst0}).} With these assumptions, and decomposing $\gamma_{\mu\nu}$ as
\eqn{medec1} \gamma_{\mu\nu}=e^{2a}\hat\gamma_{\mu\nu},\quad \det\hat\gamma=-1,\eqq
the equations of motion become
\begin{eqnarray}
\label{eoma1}4(d-1)a''+2d(d-1)a^{'2}+\frac 12\tr\left(\hat\gamma^{-1}\hat\gamma'\right)^2+\phi^{'2}+2V&=&0\\
\left(e^{da}\hat\gamma^{-1}\hat\gamma'\right)'&=&0\label{gahat}\\
\label{eoma2}\left(e^{da}\phi'\right)'-e^{da}\partial_\phi V&=&0\\
\label{hamc2} 2d(d-1) a^{'2}-\frac 12\tr\left(\hat\gamma^{-1}\hat\gamma'\right)^2-\phi^{'2}+2V&=&0.
\end{eqnarray}
Eq. (\ref{gahat}) implies that $\hat\gamma_{\mu\nu}$ has the form
\beq \label{defb}\hat\gamma_{\mu\nu}=\eta_{\mu\alpha}\left(e^{b(u)M}\right)^\alpha_{\ \nu},\eqq
where $b(u)$ is a scalar, and $M^\mu_{\ \nu}$ is a traceless covariantly constant matrix. The bulk profile of $\hat\gamma_{\mu\nu}$ is then described by a single scalar field $b$. 
Plugging (\ref{medec1}) and (\ref{defb}) subject to our restrictions on $x^\mu$-dependence of various fields in (\ref{initialaction}) gives, up to boundary terms,
\eqn{act2} S=\frac 12 \int_M d^{d+1}x e^{da}\left(2d(d-1) a^{'2}-2b^{'2}-\phi^{'2}-2V\right).\eqq
The equation of motion obtained from the above action correspond precisely to (\ref{eoma1})-(\ref{eoma2}) upon substitution (\ref{defb}), whereas (\ref{hamc2}) corresponds to the vanishing of the Hamiltonian of (\ref{act2}), i.e.
\eqn{hamc2a} 2d(d-1) a^{'2}-2b^{'2}-\phi^{'2}+2V=0.\eqq
Of course the equation of motion of $b$ is that of a free scalar, $\left(e^{da}b'\right)'=0$, so it would naively appear that one can immediately integrate out $b$. However, after integrating out $b$ condition (\ref{hamc2a}) cannot be seen as the condition of vanishing Hamiltonian of the action resulting from integrating out $b$, so it is convenient to keep $b$ off-shell.

The advantage of this structure is that now we can apply the method of fake supergravity to the action (\ref{act2}) \cite 3. Writing (\ref{act2}) \`a la Bogomol'nyi \cite{1} gives
\eqn{bpsaction} \begin{split}S_{\mathrm{eff}}=&\frac 12 \int dr e^{da}\left[2d(d-1)\left(a' +\frac 12 W\right)^2-2\left(b'-\frac{d-1}2W_b\right)^2-(\phi'-(d-1)W_\phi)^2\right]\\
&-(d-1)e^{da}W\bigg|_{\p M},\end{split}\eqq
where $W_b=\partial_b W$, $W_\phi=\partial_\phi W$, and $W$ satisfies the equation
\eqn{bpseq} \frac{(d-1)^2}2(2W_\phi^2+W_b^2)-d(d-1)W^2=2V.\eqq
Up to the boundary term, the action (\ref{bpsaction}) is a linear combination of squares, so it is extremized by
\begin{align}
\label{eqnphi1} \phi'=&\,(d-1)W_\phi\\
\label{eqnb1} b'=&\,\frac{d-1}2W_b\\
\label{eqna1} a'=&\,- \frac 12W,
\end{align}
It is easy to see that solutions to eqs. (\ref{eqnphi1})-(\ref{eqna1}) are also solutions to (\ref{eoma1})-(\ref{eoma2}) and (\ref{hamc2a}). It has actually been conjectured that the solutions of the above equations obtained for all the possible $W$ that satisfy (\ref{bpseq}) for a given $V$, are the same as the solutions to the equations of motion of (\ref{act2}) satisfying (\ref{hamc2a}) \cite{2,3}. The investigation of solutions found with the above method will be left to future work, although we will mention a simple example here. Consider $W=W(\phi)$ with $W_b=0$. Then eq. (\ref{eqnb1}) gives $b'=0$, and the equations become identical to those of one scalar coupled to a domain wall. The condition $b'=0$ corresponds to $\tr \left(\hat\gamma^{-1}\hat\gamma'\right)^2=0$, which can be shown to be equivalent to having a null deformation of the domain wall, such as those discussed in Section \ref{ndsec}. This means that any known domain wall solution, like the ones studied in \cite{14}, can be immediately extended to its null deformations.

\section{Discussion}
Our main result was a classification of the gravity solutions such that the background metric of the dual QFT is Ricci-flat, and the stress tensor is covariantly constant.

The most interesting solutions are three null deformations of AdS and three null deformations of the AdS soliton. These metrics do not suffer from scalar curvature singularities. Two null deformations of AdS, i.e. the space-times AdS II$^-$ and AdS III, appear to have no tidal force singularities. This is in contrast with AdS II$^+$, and with all the null deformations of the AdS soliton, i.e. Sol II$^+$, Sol II$^-$ and Sol III, in which there are geodesics that experience a divergent tidal force towards the IR.

It would be interesting to explore the properties of the states of the dual QFT associated to these solutions. Some work in this direction has already been pursued, for example the behavior of the entanglement entropy and the mutual information for strip shaped regions has been studied in the case of AdS II$^+$, showing interesting features \cite 7.

In Section 5 we generalized the resolution technique that we used to solve Einstein's equation, coupling the metric to a scalar. Substituting the unit determinant part of the metric with an effective scalar degree of freedom, and simplifying the system further by introducing a fictitious superpotential allows to find nontrivial solutions. These can holographically be interpreted as RG flows that violate Lorentz invariance. While it is easy to generate new solutions, it would be interesting to find some physically meaningful example of such non-Lorentz-invariant RG flows.

\vspace{0.2in}   \centerline{\bf{Acknowledgements}} \vspace{0.2in}
I would like to express my gratitude to Hong Liu for the initial motivation and for the help throughout this project. I would also like to thank Allan Adams and Julian Sonner for their helpful insights.

\appendix

\section{Resolution of Einstein's equations}\label{resalb}
Consider a $(d+1)$-dimensional space-time $M$ with a time-like hypersurface $\Sigma$.  We shall solve the Einstein's equation\footnote{The resolution presented in this Appendix can be considered as a slight generalization of the derivation given in \cite{18,18b}.}
\eqn{einstein0} E_{MN}\equiv R_{MN}-\frac 12 G_{MN}R+\Lambda G_{MN}=0,\eqq
where $G_{MN}$ is the metric on $M$ and $R_{MN}$ is the Ricci tensor. The boundary conditions are
\beq G_{MN}\partial_\mu x^M\partial_\nu x^N=h_{\mu\nu}\quad\text{on }\Sigma,\eqq
i.e. $h_{\mu\nu}$ is the induced metric on $\Sigma$, and
\beq K_{MN}\partial_\mu x^M\partial_\nu x^N=\Theta_{\mu\nu}\quad\text{on }\Sigma,\eqq
where $x^M$ is a coordinate system on $M$, $x^\mu$ is a coordinate system on $\Sigma$, $K_{MN}$ is the extrinsic curvature of $\Sigma$, i.e.
\beq K_{MN}=\frac 12 \mathcal L_n G_{MN},\eqq
where $n^M$ is the unit normal of $\Sigma$, and $h_{\mu\nu}$ and $\Theta_{\mu\nu}$ satisfy
\begin{eqnarray}
\label{ricciflat0} R_{\mu\nu}[h]&=&0\\\
\label{covconst0}D_\mu \Theta_{\nu\rho}&=&0\\
\label{hamconst1}\Theta^\mu_{\ \nu}\Theta^\nu_{\ \mu}-(\Theta^\mu_{\ \mu})^2&=&2\Lambda,\end{eqnarray}
where $R_{\mu\nu}[h]$ denotes the Ricci tensor of $h_{\mu\nu}$, and $D_\mu$ is the covariant derivative associated with $h_{\mu\nu}$. The first two conditions (\ref{ricciflat0}) and (\ref{covconst0}) will allow us to find the solution to (\ref{einstein0}) in closed form, and the last condition is necessary for the consistency with the Hamiltonian constraint \cite{11}. In a neighborhood of $\Sigma$ we can write the metric of $M$ in Gaussian normal coordinates, i.e.
\eqn{gnc1} ds^2=G_{MN}dx^M dx^N= du^2+\gamma_{\mu\nu}dx^\mu dx^\nu,\eqq
where $u$ is the affine parameter that parameterizes geodesics emanated from $\Sigma$ and orthogonal to it. In this coordinate system, $\Sigma$ is at a constant value of $u$ which we call $u_\Sigma$. Eq. (\ref{einstein0}) splits into three sectors:
\begin{eqnarray}
\label{dyn1} E^\mu_{\ \nu}&\to&\quad-\mathcal K'+ K'-K\mathcal K+\frac 12(\tr \mathcal K^2+K^2)-\frac 12 \mathrm{R}[\gamma]+\mathrm{Ric}[\gamma]+\Lambda=0,\\
\label{momconstr} E^u_{\ \mu}&\to&\quad\nabla_\alpha \mathcal K^\alpha_{\ \mu}-\nabla_\mu \mathcal K^\alpha_{\ \alpha}=0,\\
\label{ham1} E^u_{\ u}&\to&\quad-\frac 12 \tr \mathcal K^2+\frac 12 K^2-\frac 12 \mathrm{R}[\gamma]+\Lambda=0,\end{eqnarray}
where $\nabla_\mu$ denotes the covariant derivative associated to $\gamma_{\mu\nu}$, $\mathcal K$ is a matrix that denotes the extrinsic curvature tensor with mixed indices, i.e.
\eqn{extrc1}\mathcal K^\mu_{\ \nu}=K^\mu_{\ \nu}=\frac 12 \gamma^{\mu\alpha}\partial_u\gamma_{\alpha\nu},\eqq
and $K=K^\mu_{\ \mu}$. $\text{Ric}[\gamma]^\mu_{\ \nu}$ is the Ricci tensor of $\gamma_{\alpha\beta}$, and $R=R^\mu_{\ \mu}$ is the corresponding Ricci scalar. Taking
\eqn{ricciflatu} \mathrm{Ric}[\gamma]=0\eqq
for all values of the coordinate $u$, we will now work out the solution to (\ref{dyn1})-(\ref{ham1}) with boundary conditions (\ref{ricciflat0})-(\ref{hamconst1}). Assuming uniqueness of the boundary value problem, the solution we find is then the most general one with such boundary conditions. Using (\ref{ricciflatu}) and setting $\Lambda=-\frac{d(d-1)}2$, eq. (\ref{dyn1}) becomes
\eqn{dyn2}-\mathcal K'+ K'-K\mathcal K+\frac 12(\tr \mathcal K^2+K^2)-\frac{d(d-1)}2=0,\eqq
which is a first order ordinary matrix differential equation, and (\ref{ham1}) becomes
\eqn{ham2} -\frac 12 \tr \mathcal K^2+\frac 12 K^2-\frac{d(d-1)}2=0.\eqq
Eq. (\ref{dyn2}) can be split into trace and traceless parts,
\eqn{dynscal}K'+\frac 12 K^2+\frac d{2(d-1)}\tr\hat {\mathcal K}^2-\frac 12 d^2=0,\eqq
\eqn{trless}\hat{\mathcal K}'+K\hat{\mathcal K}=0,\eqq
where $\hat{\mathcal K}=\mathcal K-\frac 1d\tr\mathcal K$. Taking the derivative of (\ref{dynscal}), and using (\ref{trless}), we obtain
\eqn{dynscal1} K''+3K'K+K\left(K^2-d^2\right)=0.\eqq
The solution to (\ref{dynscal1}) is
\beq K=d\frac{c_1 c_2 e^{2d u}-1}{(1+c_1 e^{d u})(1+c_2 e^{d u})},\eqq
plugging it in (\ref{trless}), we obtain
\beq \hat{\mathcal K}=\frac{e^{d u}B}{(1+c_1e^{d u})(1+c_2 e^{d u})},\eqq
where $B$ is a traceless matrix. Plugging the above solutions into (\ref{dynscal}), we also obtain the constraint
\beq \tr B^2=d(d-1)(c_1-c_2)^2,\eqq
and plugging it into (\ref{ham1}) we get
\beq c_1+c_2=0,\eqq
and the solution then becomes
\eqn{solk1} \mathcal K=-\frac{c^2e^{2d u}+1}{1-c^2 e^{2d u}}+\frac{e^{d u}B}{1-c^2e^{2d u}},\eqq
with
\eqn{constraints1} \tr B=0,\quad \tr B^2=4d(d-1)c^2,\eqq
where we defined $c=c_1$. From (\ref{covconst0}) and the second equation in (\ref{constraints1}) we infer that $c$ is a constant. From this fact, from (\ref{covconst0}) and from (\ref{solk1}),
\beq \nabla_\mu B^\alpha_{\ \beta}=0,\eqq
which then implies that
\beq \nabla_\mu K^\alpha_{\ \beta}=0,\eqq
everywhere, and therefore (\ref{momconstr}) is automatically satisfied. Obviously (\ref{hamconst1}) is satisfied as it is just (\ref{ham2}) evaluated at $u_\Sigma$. To obtain the metric we integrate eq. (\ref{extrc1})
\beq \frac 12 g^{-1}g'=\mathcal K,\eqq
which has solution
\beq g=he^{2\int_{u_\Sigma}^u ds \mathcal K}=he^{2(u_\Sigma-u)}\left(\frac{1-c^2e^{2d u}}{1-c^2 e^{2d u_\Sigma}}\right)^{\frac 2d}\left(\frac{1+ce^{d u}}{1-ce^{d u}}\frac{1-c e^{d u_\Sigma}}{1+ce^{d u_\Sigma}}\right)^{\frac B{d c}},\eqq
where we imposed that $g=h$ at $u=u_\Sigma$. To recover (\ref{sol0}), we take $u_\Sigma\to -\infty$ with
\eqn{tb1}g_{\mu\nu}= e^{2u_\Sigma}h_{\mu\nu},\quad T^\mu_{\ \nu}=2 B^\mu_{\ \nu},\quad u_0=-\frac 1d\log c,\eqq
where $g_{\mu\nu},\ T^\mu_{\ \nu}$ and $u_0$ are taken to be independent of $u_\Sigma$. The solution then becomes
\beq \gamma=ge^{-2u}\left(1+ e^{d(u-u_0)}\right)^{\frac 2d (1+2^{-2}e^{du_0}T)}\left(1-e^{d(u-u_0)}\right)^{\frac 2d (1-2^{-2}e^{du_0}T)},\eqq
and (\ref{constraints1}) implies
\beq\tr T=0,\quad \tr T^2=16d(d-1)e^{-2du_0}.\eqq

\section{Residual symmetries of the metric}\label{appendixa}
Consider a metric in the coordinates (\ref{gnc11}) that depends only on $u$ and with boundary condition (\ref{guv1}). To preserve (\ref{gnc11}), an infinitesimal coordinate transformation $\xi^P$ should satisfy
\eqn{b1q}\mathcal L_\xi G_{Mu}=\xi^P \partial_P G_{M u}+\partial_M\xi^PG_{Pu}+\partial_u\xi^PG_{MP}=0.\eqq
The solution to the above equation is
\eqn{trivial} \xi^u=f(x),\qquad \xi^\mu=C^\mu(x)-\partial_\nu f(x)\int_0^u G^{\mu\nu}(r,x) dr,\end{equation}
and the corresponding variation of $G_{\mu\nu}$ at $u=0$ is
\beq \mathcal L_\xi G_{\mu\nu}=\xi^u \partial_uG_{\mu\nu}=f(x)\partial_uG_{\mu\nu}.\eqq
Since we want the metric to remain independent of the boundary coordinates, we require $\partial_\mu f(x)=0$. We also want the transformations to preserve (\ref{guv1}), i.e. $g_{\mu\nu}=\eta_{\mu\nu}$. We then conclude that the subgroup of (\ref{trivial}) that satisfies the three requirements mentioned above (\ref{b1q}) is given by the Poincar\`e transformations acting on $x^\mu$.

\section{Scalar curvature singularities}\label{curvsing}
We call scalar curvature singularities points of space-time in which a scalar quantity constructed from the Riemann tensor and its derivatives, such as $R^{MNAB}R_{MNAB}$, is divergent. In this Appendix we study such singularities for the solutions discussed in Section \ref{classification}. In the coordinate system (\ref{gnc11}) the Riemann tensor can be decomposed as
\begin{eqnarray}
\label{ruu} R^\alpha_{\ u\beta u}&=&-K^{'\alpha}_{\ \beta}-K^\alpha_{\ \gamma} K^\gamma_{\ \beta}\\
\label{rubeta} R^\alpha_{\ u\beta\gamma}&=&\nabla_\beta K^\alpha_{\ \gamma}-\nabla_\gamma K^\alpha_{\ \beta}\\
\label{rmunu} R^\mu_{\ \nu\sigma\tau}&=&K^\mu_{\ \tau} K_{\sigma\nu}-K^\mu_{\ \sigma} K_{\tau\nu}+\tilde R^\mu_{\ \nu\sigma\tau},\end{eqnarray}
where $\nabla_\beta$ is the covariant derivative associated with $\gamma_{\alpha\beta}$, and $\tilde R^\mu_{\ \nu\sigma\tau}$ is the intrinsic Riemann tensor on hypersurfaces with constant $u$. For the homogeneous solutions of Section \ref{classification} the RHS of (\ref{rubeta}) vanishes, as well as $\tilde R^\mu_{\nu\sigma\tau}$. Consider first type I and type IV solutions, whose boundary stress tensor $T^\mu_{\ \nu}$ can be put in diagonal form (and can possibly have complex eigenvalues). According to our analysis, such solutions are either of the form (\ref{sol0}) with $u_0$ finite, or are of the form (\ref{sol0a}) with $T=0$. The latter case gives only pure AdS and there are no singularities. In the former case most of the metrics have scalar curvature singularities at $u_0$. To see this, consider the expansion of the extrinsic curvature of (\ref{sol0}) near $u_0$:
\beq K^\mu_{\ \nu}=\frac 1{d(u_0-u)}\left(\delta^\mu_\nu-\frac 14e^{du_0}T^\mu_{\ \nu}\right).\eqq
We now assume that the above expression is diagonal, i.e. $T^\mu_{\ \nu}=t_{(\mu)}\delta^\mu_\nu$. From (\ref{ruu})-(\ref{rmunu}), the Riemann tensor in this region is
\begin{eqnarray} R^{\mu u}_{\ \ \mu u}&=&-\frac 1{(d(u_0-u))^2}\left(1-\frac 14e^{d u_0}t_{(\mu)}\right)\left(1-d-\frac 14e^{d u_0}t_{(\mu)}\right)\\
R^{\mu\nu}_{\ \ \mu\nu}&=&-\frac 1{(d(u_0-u))^2}\left(1-\frac 14e^{d u_0}t_{(\mu)}\right)\left(1-\frac 14e^{d u_0}t_{(\mu)}\right),\end{eqnarray}
where there is no sum on the indices, and the other components are zero (unless related to the above by exchange symmetries). Consider now the tensor
\beq S^A_{\ E}=R^{A B}_{\ \ CD}R^{CD}_{\ \ E B},\eqq
we have $S^u_{\ \mu}=S^\mu_{\ u}=0$, and
\begin{eqnarray} S^u_{\ u}&=&2\sum_\mu (R^{\mu u}_{\ \ \mu u})^2=\frac 2{(d(u_0-u))^2} \sum_\mu \left(1-\frac14e^{d u_0}t_{(\mu)}\right)\left(1-d-\frac14e^{d u_0}t_{(\mu)}\right)\label{esuu}\\
S^\mu_{\ \nu}&=&2\delta^\mu_\nu\left(\sum_{\beta\neq \mu}(R^{\mu\beta}_{\ \ \mu\beta})^2+(R^{\mu u}_{\ \ \mu u})^2\right)\\
&=&\frac 4{d(u_0-u)^2}\delta^\mu_\nu\left(1-\frac14e^{d u_0}t_{(\mu)}\right)\left(d-1+\frac14e^{d u_0}t_{(\mu)}\right)\qquad\quad\label{esmunu}\end{eqnarray}
where in the last passage of the last equation we used (\ref{sol10}) and (\ref{sol20}). The trace of any power of $S^A_{\ B}$ is a curvature invariant of space-time and, in order to have none of such traces to diverge, each of the eigenvalues of $S^A_{\ B}$ has to be finite. From the expressions in (\ref{esuu}) and (\ref{esmunu}) we then infer that $t_{(\mu)}$ have to be equal to either $4e^{-du_0}$ or $-4(d-1)e^{-du_0}$. In order to be consistent with (\ref{sol10}) and (\ref{sol20}), the only possibility is that all but one among $t_{(\mu)}$ be equal to $4e^{-du_0}$, and the remaining one be equal to $-4(d-1)e^{-du_0}$. For $d>2$, this rules out most of the solutions of type I and all the solutions of type IV.

We now extend the above analysis to solutions of type II and type III. From (\ref{solk1}) and the second eq. in (\ref{tb1}), and from the discussion around (\ref{t23}), the extrinsic curvature can be written in the form
\beq K^\mu_{\ \nu}=f_1(u)\delta^\mu_\nu+f_2(u)D^\mu_{\ \nu}+f_3(u)H^{\mu}_{\ \nu},\eqq
where $D^\mu_{\ \nu}$ is diagonalizable, $H^\mu_{\ \nu}$ is nilpotent and $[D,H]=0$. Taking into account that all the fields depend only on the $u$ coordinate, from (\ref{ruu})-(\ref{rmunu}) it is easy to see that any curvature scalar can be written in terms of traces of powers of $K^\mu_{\ \nu}$ and its $u$-derivatives. Now,
\beq \tr \left[(\partial_u^{(n_1)} K)^{m_1}\cdots (\partial_u^{(n_N)} K)^{m_N}\right]= \tr \left[(\partial_u^{(n_1)}D)^{m_1}\cdots (\partial_u^{(n_N)}D)^{m_N}\right],\eqq
where $(\partial_u^{(n_i)} K)^{m_i}$ denotes the $m_i$-th power of the $n_i$-th derivative of $ K^\mu_{\ \nu}$ with respect to $u$. In the above equality we used the fact that $(\partial_u^{(n_i)}f(u) H)^{m_i}=(\partial_u^{(n_i)}f(u))^{m_i}H^{m_i}$ is nilpotent, and that $\tr(MN)=0$ if $N$ is nilpotent and $M$ is a matrix such that $[M,N]=0$. The above equation implies that the curvature scalars depend only on the diagonal part of the extrinsic curvature $D^\mu_{\ \nu}$. We can then readily extend the conclusion reached for type I and type IV to type II and type III metrics, i.e. there are no scalar curvature singularities if an only if all but one eigenvalues of the boundary stress tensor $T^\mu_{\ \nu}$ are equal to $4 e^{-du_0}$ and the leftover eigenvalue is equal to $-4(d-1)e^{-du_0}$.

\section{Tidal force singularities}\label{apptidal}
The solutions listed in Table \ref{table1} have no scalar curvature singularities, however some of their geodesics experience a divergent amount of tidal force as they approach the IR region. Consider a geodesic with tangent vector $v^M(\lambda)=(\dot u(\lambda),\dot x^\mu(\lambda))$, where $\lambda$ is an affine parameter and the dot denotes the derivative with respect to $\lambda$. For a unit vector $\eta^M$ orthogonal to $v^M$ the tidal force is described by the quantity
\eqn{tide1} v^M \nabla_M(v^N\nabla_N \eta^A).\eqq 
In this Appendix we shall study for which geodesics the above expression diverges. The metrics given in Table \ref{table1} are invariant under translations along the boundary coordinates, $x^\mu\to x^\mu+a^\mu$, where $a^\mu$ is a constant $d$-dimensional vector. This implies that $v_{\mu}$ is constant along the geodesic, i.e. $p_0\equiv v_0=const.$, $p_i\equiv v_i=const.$. These equations, together with the normalization condition $v^M v_M=\varepsilon$, where $\varepsilon=0,\pm1$, allow us to easily find all the geodesics of the metrics we want to study.

The space-times AdS II$^+$ and AdS II$^-$ given in (\ref{nametric}) have geodesics generated by the velocity vector
\begin{eqnarray}\label{geod01}
\dot u^2&=&e^{2u}(p_0^2-p_1^2-p_\perp^2)+\frac 1d\gamma e^{(d+2)u}(p_0+p_1)^2+\varepsilon\\
\dot x^0&=&-e^{2u}p_0-\frac 1d \gamma e^{(d+2)u}(p_0+p_1)\\
\label{geod02} \dot x^1&=&e^{2u}p_1-\frac 1d \gamma e^{(d+2)u}(p_0+p_1),\quad \dot x^\perp=e^{2u}p_\perp,\end{eqnarray}
where $\gamma=+1$ for AdS II$^+$ and $\gamma=-1$ for AdS II$^-$. For $\gamma=-1$ we see that in general $\dot u^2$ changes sign when the geodesic approaches the deep IR $u\to\infty$, which corresponds to a turning point. It is easy to see that the only way to avoid the change of sign of $\dot u^2$ is to require $p_0+p_1=p_\perp^2=0$, which implies $\dot u^2=const.$. There are thus no geodesics that can reach $u=\infty$ in a finite proper time, and since all the other points of AdS II$^-$ are regular, we conclude that geodesics do not detect tidal force singularities, although an arbitrary high tidal force can be detected if a geodesic goes sufficiently deeply in the IR (which can be achieved by taking $p_0+p_1$ slightly positive, and $p_0-p_1$ positive and large). This situation is in contrast with that of AdS II$^+$ as, when $\gamma=1$, $\dot u^2$ remains in general positive as $u\to\infty$, and geodesics can reach $u=\infty$ in a finite proper time. Choosing the unit vector $\eta^M=(\eta^u,\eta^\mu)=(0,0,\dots,e^u)$, the quantity (\ref{tide1}) for a geodesic of the form (\ref{geod01}),(\ref{geod02}) gives
\beq v^M \nabla_M(v^N\nabla_N \eta^A)=\left(-2^{-1}e^{(2+d)u}(p_0+p_1)^2+\varepsilon\right)\eta^A,\eqq
when $d>2$, which diverges as $u\to\infty$, and thus in this case we do have a tidal force singularity at $u=\infty$.

The metrics Sol II$^+$ and Sol II$^-$ are given in (\ref{solit2}), and their geodesics are generated by the velocity vector
\begin{eqnarray}\label{geods01} \dot u^2&=&\left(2\cosh\frac{du}2\right)^{-\frac 4d}\left(p_0^2-p_1^2-p_\perp^2+\frac 1d\gamma (p_0+p_1)^2\arctanh\left(e^{du}\right)\right)+\varepsilon\\
\dot x^0&=&\left(2\cosh\frac{du}2\right)^{-\frac 4d}\left(-p_0-\frac 1d \gamma(p_0+p_1) \arctanh\left(e^{du}\right)\right),\\
\dot x^1&=&\left(2\cosh\frac{du}2\right)^{-\frac 4d}\left(p_1-\frac 1d \gamma (p_0+p_1)\arctanh\left(e^{du}\right)\right),\label{geods02}\\
\dot x^2&=&0,\quad \dot x^\perp=\left(2\cosh\frac{du}2\right)^{-\frac 4d}p_\perp,\end{eqnarray}
where for simplicity we set $u_0=0$ and also $p_2=0$, as the latter gives a repulsion term, which near $u=0$ goes like $u^{-2}$, and it would prevent the geodesic to reach the singularity, which we want to study. This repulsion term is also present in the AdS soliton and in that case it can be thought of as the centrifugal force due to the compactification of the $x^2$ coordinate. For $\gamma=-1$ the geodesics reach $u=0$ only if $p_0+p_1=0$, in which case the geodesics become, near $u=0$,
\beq \dot u^2=\varepsilon-2^{-\frac 4d}p_\perp^2,\quad \dot x^0=\dot x^1=-2^{-\frac 4d}p_0,\quad \dot x^2=0,\quad \dot x^\perp=2^{-\frac 4d}p_\perp,\eqq
which shows that $u=0$ can be reached by space-like geodesics if $p_\perp^2$ is small enough, and it takes a finite proper time. Choosing $p_0=p_\perp=0$ for simplicity, and $\eta^M$ near $u=0$ to behave as
\beq \eta^M=\left(0,1,-1,\frac{2\sqrt{-\log(-u)}}{\sqrt d},0\dots,0\right),\eqq
and we find, as $u\to 0$,
\beq v^M \nabla_M(v^N\nabla_N \eta^A)=\frac 1{2du^2}(0,1,1,0,\dots,0),\eqq
we thus conclude that space-like geodesics of Sol II$^-$ can detect tidal force singularity at $u=0$. For $\gamma=1$ it can be easily seen from (\ref{geods01})-(\ref{geods02}) that it is possible to reach $u=0$ with time-like geodesics in a finite proper time. Setting e.g. $p_0=p_1$ and $p_\perp=0$, with $\varepsilon=-1$, and choosing $\eta^M$ near $u=0$ to behave as $\eta^M=\left(0,0,0,4^{-\frac 1d},0,\dots,0\right)$, it can be found that, as $u\to 0$,
\beq v^M \nabla_M(v^N\nabla_N \eta^A)=\left(0,0,0,-2^{1-\frac 6d}(d+2)p_0^2\log(-u),0,\dots,0\right),\eqq
and we conclude that Sol II$^+$ has a tidal force singularity at $u=0$, which can be detected by time-like geodesics.

For the AdS III metric (\ref{ads3}), the geodesics are generated by the velocity vector
\begin{eqnarray}\dot u^2 &=&e^{2u}(p_0^2-p_1^2-p_2^2-p_\perp^2)+\frac 2de^{(d+2)u}p_2(p_0+p_1)-\frac{1}{2d^2}e^{2(d+1)u}(p_0+p_1)^2+\varepsilon\\
\dot x^0&=&-e^{2u}p_0-\frac 1de^{(d+2)u}p_2+\frac{1}{2d^2}e^{2(d+1)u}(p_0+p_1)\\
\dot x^1&=&e^{2u}p_1-\frac 1de^{(d+2)u}p_2+\frac {1}{2d^2}e^{2(d+1)u}(p_0+p_1)\\
\dot x^2&=&e^{2u}p_2-\frac1d e^{(d+2)u}(p_0+p_1),\quad \dot x^\perp=e^{2u}p_\perp.
\end{eqnarray}
Similarly to AdS II$^-$, almost all the geodesics $\dot u^2$ will change sign as approaching $u=\infty$, which corresponds to a turning point of the geodesic. To avoid this turning point it is necessary to have $p_0+p_1=p_2^2=p_\perp^2=0$, in which case we are left with the geodesics generated by
\beq \dot u=1,\quad \dot x^0=\dot x^1=-e^{2u}p_0,\quad \dot x^2=\dot x^\perp=0,\eqq
which can only be space-like, and take an infinite proper time to reach the IR. The situation is then the same as that of AdS II$^-$, and we conclude that no geodesic can detect tidal force singularities.

The metric Sol III (\ref{solit3}) gives geodesics generated by
\begin{eqnarray}\dot u^2&=&\left(2\cosh\frac{du}2\right)^{-\frac 4d}\bigg(p_0^2-p_1^2-p_2^2-p_\perp^2+\frac 2dp_2(p_0+p_1)\arctanh\left(e^{du}\right)\\
&&-\frac 1{2d^2}(p_0+p_1)^2\arctanh^2\left(e^{du}\right)\bigg)+\varepsilon\\
\dot x^0&=&\left(2\cosh\frac{du}2\right)^{-\frac 4d}\left(-p_0-\frac 1dp_2\arctanh\left(e^{du}\right)+\frac 1{d^2}(p_0+p_1)\arctanh^2\left(e^{du}\right)\right)\\
\dot x^1&=&\left(2\cosh\frac{du}2\right)^{-\frac 4d}\left(p_1-\frac 1dp_2\arctanh\left(e^{du}\right)+\frac 1{2d^2}(p_0+p_1)\arctanh^2\left(e^{du}\right)\right)\\
\dot x^2&=&\left(2\cosh\frac{du}2\right)^{-\frac 4d}\left(p_2-\frac 1{d}(p_0+p_1)\arctanh\left(e^{du}\right)\right)\\
\dot x_\perp&=&\left(2\cosh\frac{du}2\right)^{-\frac 4d}p_\perp,
\end{eqnarray}
where we set $u_0=p_3=0$ for simplicity. In particular, $p_3$ multiplies a term proportional to $u^{-2}$ in the expression of $\dot u^2$, which has the same repulsive effect as turning on $p_2$ in Sol II$^+$ and Sol II$^-$. Before reaching $u=0$, $\dot u^2$ flips sign and we thus have a turning point. The only way to avoid this case is again to take $p_0+p_1=0$. In this case the geodesic can only be space-like. We choose the simplest case, with $\dot u=1$ and $\dot x^\mu=0$, and the hypersurface $u=0$ can be reached in a finite proper time. We look for tidal force singularities by taking $\eta^M$ such that, near $u=0$, $\eta^M=(0,0,0,2^{-\frac 2d},0,0,\dots,0)$,
which then gives
\beq v^M \nabla_M(v^N\nabla_N \eta^A)=\frac 1{du^2} 2^{-1-\frac 2d}(0,1,1,0,\dots,0),\eqq
and we conclude that Sol III has a tidal force singularity at $u=0$, which can be detected by space-like geodesics. The structure is then similar to that of Sol II$^-$.

\section{Holonomy groups of Ricci-flat space-times}\label{holap}
Here we briefly introduce and describe the holonomy groups that we used in Sec. \ref{nonhomsec}. For simplicity, we restrict the discussion to simply connected 4-dimensional space-times. Since parallel transport preserves the inner product between two vectors, the holonomy group of a simply connected space-time $V$ is a connected subgroup of the Lorentz group. As shown in \cite 5, such subgroups are in one-to-one correspondence with the Lie subalgebras of the Lorentz algebra $L$, which is the set of antisymmetric rank-2 tensors $A_{\mu\nu}=-A_{\nu\mu}$. The algebra $L$ can be generated by the antisymmetric tensor products \cite 5
\eqn{genlor} l\wedge n,\quad l\wedge x,\quad l\wedge y,\quad n\wedge x,\quad n\wedge y,\quad x\wedge y,\eqq
where $(l,n,x,y)$ is a null basis for the space-time, i.e. the only nonvanishing products between the vectors are $l\cdot n=x^2=y^2=1$, and
where, for example $(l\wedge x)_\mu^{\ \nu}=l_\mu x^\nu-x_\mu l^\nu$. The 6-dimensional Lorentz algebra $L$ admits fifteen types of Lie subalgebras which, up to isomorphisms, can be spanned by combinations of the above antisymmetric products. For the above discussion, these fifteen subalgebras provide a classification of the holonomy groups of simply connected 4-dimensional space-times. In the case of Ricci-flat space-times, the Riemann tensor has additional algebraic properties which restrict the classification to four possible holonomy groups. In the notation of \cite 5 they are denoted by $R_1,R_8,R_{14}$ and $R_{15}$. $R_1$ is the identity, i.e. it corresponds to trivial holonomy, and is the holonomy group of flat space-times. The Lie algebra of $R_8$ is generated by
\beq l\wedge x,\quad l\wedge y,\eqq
and the Lie algebra of $R_{14}$ is generated by
\beq l\wedge x,\quad l\wedge y,\quad l\wedge n,\quad x\wedge y.\eqq
$R_{15}$ is generated by all six generators in (\ref{genlor}), and it thus represents the whole identity connected component of the Lorentz group.

\end{document}